\documentclass[twocolumn,showpacs,preprintnumbers,amsmath,amssymb]{revtex4-1}

\usepackage{graphicx}
\usepackage{dcolumn}
\usepackage{bm}
\usepackage{float}
\usepackage{lineno}
\begin{document}

\title{Nonlinear Electronic Stopping for Slow Ion in a Narrow Band Gap Semiconductor: Formation of Chemical Bonds during Collision}



\author{Chang-Kai Li$^{1,2}$, Qiang Cao$^{1}$, Feng Wang$^{3}$, Xiao-Ping OuYang$^{4}$, {Sheng Liu}$^{1*}$ and {Feng-Shou Zhang}$^{2,5,6}$}
\email[]{Corresponding author. victor$_$liu63@vip.126.com}
\email[]{Corresponding author. fszhang@bnu.edu.cn}
\affiliation{%
$^{1}$The Institude of Technological Science, Wuhan university, Wuhan 430072, China\\
$^{2}$The Key Laboratory of Beam Technology and Material Modification of Ministry of Education, College of Nuclear Science and Technology, Beijing Normal University, Beijing 100875, China \\
$^{3}$School of Physics, Beijing Institute of Technology, Beijing 100081, China\\
$^{4}$Northwest Institute of Nuclear Technology, Xi¡¯'an 710024, China\\
$^{5}$Beijing Radiation Center, Beijing 100875, China\\
$^{6}$Center of Theoretical Nuclear Physics, National Laboratory of Heavy Ion Accelerator of Lanzhou, Lanzhou 730000, China\\
}%

\date{\today}

\begin{abstract}

Using time-dependent density-functional theory, we investigate the electronic stopping power of self-irradiated silicon through non-adiabatic dynamics simulations. For specific velocities above 0.6 atom units, electronic stopping shows a generally assumed metallic behavior that is velocity-scaling.
While in the lower velocity regime, the slope of electronic stopping power versus velocity changes and the overall magnitude are significantly greater than expectations, leading to a complete vanish of hard threshold.
An analysis of the 
electron localization function allows us to arrive at the following conclusion: the long duration of encounter process between the host atoms and the projectiles with low velocities makes possible the formation of chemical bonds with relative high bond order.
The continuous formation and breaking of chemical bonds provides an additional effective energy loss channel.


\end{abstract}


\maketitle
Due to the extreme harsh operating environment, the required prolonged operability of nuclear and space facilities 
poses an unprecedented challenge to their radiation tolerance. Radiation over certain level is known to cause irreversible deteriorations in their performance, this is particular true for the semiconductor radiation detectors \cite{FRANKS199995}.
For this reason, the effects of radiation has been a subject of extensive research. Knowing how radiation damage in materials is initiated and evolves over time is a question of fundamental importance.


In the physics of ions moving through solids, it is customary to differentiate between the
energy transferred by the projectile to nuclear motions, the rate of which is called the nuclear
stopping power \emph{$S_{n}$}, and the energy transferred to electronic excitations, giving the electronic
stopping power \emph{$S_{e}$}. The distinction is relevant since the different masses of nuclei and electrons
give rise to different response regimes: the nuclear stopping power dominates at lower projectile
velocities, while the electrons respond more readily at higher speeds.
Experimentally, it is hard to extract the electronic component for velocity regime below 0.1 a.u. \cite{PhysRevA.88.032901}, due to the sizable contribution from the nuclear stopping.  
Theoretically, it is a convenient
decoupling, since two limits can be studied and understood more easily: electronic excitations
for frozen nuclei in the high-velocity regime, and atomic motions with electrons following
adiabatically in the slow-velocity regime.

Radiation damage simulations performed to date \cite{PhysRevB.75.054106,PhysRevB.74.014101} largely rely on classical picture of ion-ion collisions, where the electrons are restricted to the Born-Oppenheimer (BO) adiabatic surface, since it is known that nuclear stopping dominates. The actual interatomic forces could be significantly altered, however, by the local electron heating produced by
the excitation of target electrons. Thus, the non-adiabatic electronic force is of equal significance not only during the initial stage but also in the cooling phase of radiation damage cascade.

Great efforts have been devoted to quantify the \emph{$S_{e}$} during ion--solid interaction.
For fast ions with kinetic energy MeV/u, Bethe and Bloch \cite{bethe,bloch} and improvements thereafter \cite{Lindhard1954ON,PhysRevLett.11.26} yielded both qualitative and quantitative agreement with the experiments. At low projectile velocity $v$ $\leq$ $v_{F}$ ($v_{F}$ denotes the Fermi velocity of the target electrons), \emph{$S_{e}$} is commonly assumed to be \emph{$S_{e}$} $\propto$ $v$ \cite{Race2010The,Vald1993Electronic,Mart1996Energy,Pitarke1999Band,PhysRevLett.118.103401}. In recent years, modeling the electronic power of ions with velocities below the Bohr velocity i.e. atom units (a.u., hereafter) is of special interest. In this regime ions are only partially stripped, quantifying the effective charge \cite{PhysRevB.86.094102,PhysRevB.94.041108,PhysRevB.94.155403,PhysRevA.61.032901} and the charge transfer \cite{PhysRevA.87.032711,PhysRevB.67.205426,PhysRevLett.119.103401,PhysRevLett.107.163201} between the projectile and host atoms is quite a challenge.

Recently, advances in high--performance computing have paved the way to follow electronic dynamics during ion--solid collision based on real-time time-dependent density functional theory (RT-TDDFT).
Besides the research of electronic excitations, the charge transfer and electronic structure, even the chemical process \cite{nature} during the collision can be handled effectively by TDDFT.  
TDDFT complements a variety of other atomistic and non-atomistic techniques with adjustable parameters.

In this letter, we propose to investigate the \emph{$S_{e}$} of self-irradiated Si under channeling condition. Glancing collisions with host atoms confine the trajectory of a channelling ion, with most of its energy dissipated through electronic excitation. Further more, to completely guard against the effect of \emph{$S_{n}$}, the host ions are frozen in the equilibrium positions. The projectiles are aligned with the middle axis of $<$100$>$ channel with a given descending velocity along the negative $z$-direction. For velocity regime above 0.1 a.u., the projectiles are free to move,
the key quantity of interest \emph{$S_{e}$} is extracted from the change in kinetic projectile energy using the thickness of the target. For velocity below 0.1 a.u., it is impossible for channeling, the projectiles are dragged with a constant velocity, and \emph{$S_{e}$} is derived from the increasing rate of total system energy. \emph{$S_{e}$} by the above two methods yields agreement within 0.7\% for velocities above 0.1 a.u..

The collision of intruding ions with the host nuclei and electrons are characterized by the Ehrenfest coupled electron--ion dynamics combined with time--dependent density--functional theory (ED--TDDFT) \cite{PhysRevLett.108.225504,Calvayrac2000Nonlinear,Alonso2008Efficient,Page2008The}.
The simulations were carried out by using the OCTOPUS \emph{ab initio} real-space code \cite{marques2003octopus,Castro2006Octopus} and employing the adiabatic local-density approximation \cite{PhysRevB.45.13244} for the
time-dependent exchange-correlation potential.
A time step of 0.001 fs is adopted to ensure the stability of the computation, simulations with smaller time steps give essentially the same results.
Furthermore, a fcc-structured 2$\times$2$\times$3 conventional cell comprising 96 Si atoms is employed with a lattice constants of 5.43 {\AA}, identical to the measured value \cite{Wyckoff2}. Halving the number of host atoms produce only slight influence to \emph{$S_{e}$} for velocity regime investigated in this work (with discrepancy within 7\% at 0.1 a.u. and 2.5\% at 1 a.u.), which indicates the finite size effect is insignificant in such problem.

\begin{figure}[htp]
  \centering
  \includegraphics[width=8.6cm,height=6.5cm]{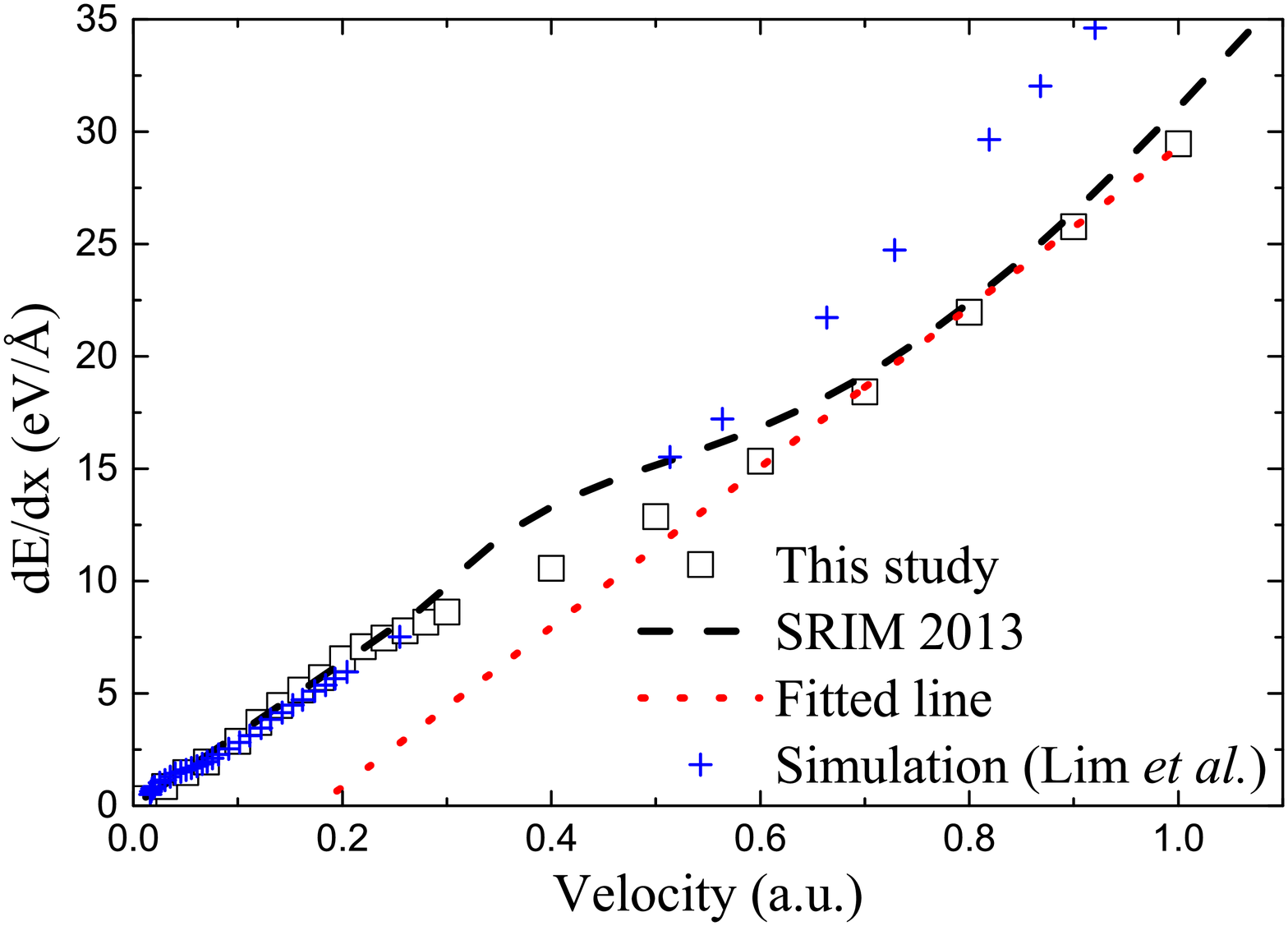}\\
  \caption{ (Color online) Electronic stopping power for silicon ions as a function of velocity along the middle axis of $<$100$>$ channel, together with the SRIM 2013 predictions and simulation results reported in Ref.~\cite{PhysRevLett.116.043201}. The SRIM data have been scaled by 2.81.}
\end{figure}

Figure 1 shows our calculated results for \emph{$S_{e}$} of silicon ions in bulk Si with velocity range 0.1 -- 1.0 a.u., together with the predictions from the SRIM--2013 database and the simulated results by Lim $et$ $al.$ \cite{PhysRevLett.116.043201}. There is a quantitative agreement between our simulated results and SRIM data. It is to be noted the SRIM data have been scaled by 2.81. As reported by the previous works, channeling generally would underestimate the \emph{$S_{e}$} due to relatively low electron density along the trajectory \cite{PhysRevLett.116.043201,PhysRevLett.108.213201}.
An interesting phenomenon we find in Fig.\,1 is that \emph{$S_{e}$} shows two distinctly different trends. Velocity-proportionality assumption is valid for higher velocity regime ($v$ $>$ 0.6 a.u.), intercepting zero at a finite velocity of 0.18 a.u.. This indicates a definitive
threshold, conforming to the traditional interpretation of electronic stopping in materials with band gap \cite{PhysRevB.91.125203,Markin2009Vanishing,PhysRevLett.84.5300,Pruneda2007Electronic}. While for the lower velocity regime, \emph{$S_{e}$} shows pronounced nonlinearities and the amplitude is significantly higher than expectations (the fitted line), leading to complete vanishing of the hard threshold. The SRIM data and the simulated results by Lim $et$ $al.$ generally show similar character.

In the present work, \emph{$S_{e}$} is found to be related with the formation and breaking of chemical bond between the projectile and host atoms, which is deemed as an additional energy dissipation channel besides electron--hole pair excitation. The intuitive concept of a chemical bond is very simple
and elegant: an electron pair shared between neighboring atoms or localization attractors that provides the necessary attraction to bind the molecule. Breaking the electron pair when the projectile leaving the neighboring atoms which is companied by the transition from the $\pi$ bonding state to $\pi^{*}$ anti-bonding state will consume additional energy.

The nonlinear phenomenon in Fig.\,1 is studied by checking time-dependent electron localization function (ELF) \cite{PhysRevA.71.010501,nature}. ELF is a robust descriptor of chemical bond based on topological analysis of local quantum mechanical functions
related to the Pauli exclusion principle. It can be deemed as the probability of finding a localized electron pairs of \emph{opposite} spins between two neighboring atoms. Its topological basins greatly resemble simple chemical pictures of where electrons, bonding and nonbonding, ought to be.
Furthermore,
the time-dependent ELF allows the time-resolved observation of the formation and breaking of chemical bonds, and can thus provide a visual understanding of complex
reactions involving the dynamics of excited electrons.

Practically, an alternative measure of
delocalization function $D_{\sigma}(\textbf{r},t)$ that defined as the probability of finding one electron
in the near vicinity of a reference like-spin electron at position $\textbf{r}$ and time $t$ is usually adopted. Based on the Pauli exclusion
principle, if $D_{\sigma}(\textbf{r},t)$ is high, the reference electron must be delocalized. For a determinantal many-body wave
built from Hartree-Fock or Kohn-Sham orbitals, $D_{\sigma}(\textbf{r},t)$ is defined as
\begin{eqnarray}
D_{\sigma}(\textbf{r},t)=\tau_{\sigma}(\textbf{r},t)-\frac{1}{4} \frac{[\nabla n_{\sigma}(\textbf{r},t)]^{2}}{n_{\sigma}(\textbf{r},t)}-\frac{j_{\sigma}^{2}(\textbf{r},t)}{n_{\sigma}(\textbf{r},t)},
\end{eqnarray}
where $\sigma$ denotes the spin, $n_{\sigma}$ the spin density, $j_{\sigma}$ the absolute value of the current density,
and
\begin{eqnarray}
\tau_{\sigma}(\textbf{r},t)=\sum^{N_{\sigma}}_{i=1}|\nabla\varphi_{i\sigma}(\textbf{r},t)|^{2},
\end{eqnarray}
the kinetic-energy density of
a system of $N_{\sigma}$
electrons, described by the single-particle orbitals $\varphi_{i\sigma}$.
The ELF can then be given by
\begin{eqnarray}
ELF(\textbf{r},t)=\frac{1}{1+[D_{\sigma}(\textbf{r},t)/D_{\sigma}^{0}(\textbf{r},t)]^{2}},
\end{eqnarray}
with the definition of
\begin{eqnarray}
D_{\sigma}^{0}(\textbf{r},t)= \tau_{\sigma}^{HEG}(n_{\sigma}(\textbf{r},t)),
\end{eqnarray}
where
\begin{eqnarray}
\tau_{\sigma}^{HEG}(n_{\sigma})=\frac{3}{5}(6\pi^{2})^{2/3}n_{\sigma}^{5/3},
\end{eqnarray}
is the kinetic-energy density of a homogeneous electron gas with spin density equal to the local value of $n_{\sigma}$. The ELF is thus
a dimensionless localization index calibrated with respect to the HEG as reference.
According to Eq.~3, there is a reverse relationship between the ELF and $D_{\sigma}$, and the values of ELF are restricted to the range 0 $\leq$ ELF $\leq$ 1. In a homogeneous electron gas, ELF has everywhere the value 1/2.

\begin{figure}[htp]
  \centering
  \includegraphics[width=8.6cm,height=8cm]{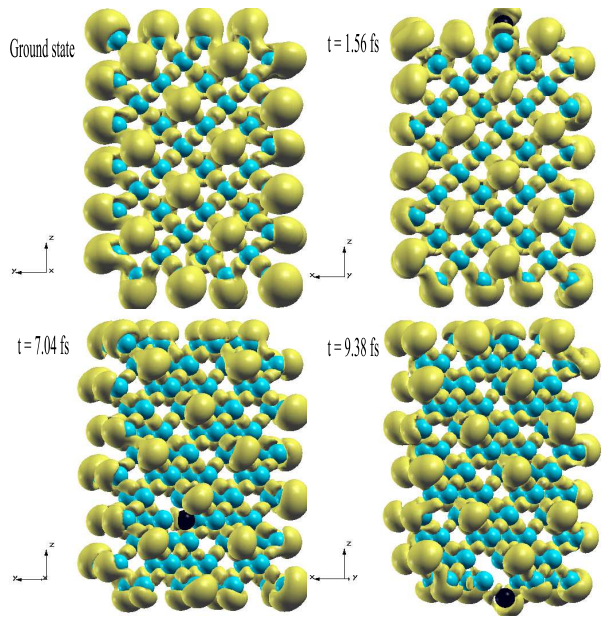}\\
  \caption{(Color online) Snapshots of time evolution of the ELF induced by a Si$^{4+}$ ion moving through bulk Si crystal along $<$100$>$ direction with $v$ = 0.1 a.u. (side view). The black ball is the projectile, the blue ones are the host atoms, the gray region is ELF isosurface. The isosurface is plotted at ELF = 0.8.}
\end{figure}

As a first step, the time evolution of a silicon ion moving through the $<$100$>$ channel for a given velocity of 0.1 a.u. is visualized in Fig.\,2. Four snapshots covering the entire collision process are
presented. In the beginning of the simulation, the crystal is in its ground state. At this moment, the
ELF exhibits two major features: four tori between the silicon atoms inside the crystal and its neighboring atoms -- the silicon-silicon single bond, and the characteristic blobs around the silicon atoms on the edge of the crystal -- lone pairs that do not form chemical bond. In the following penetrating process, the tori between the projectile and neighboring host silicon atoms are clear visible,
which is deemed as an qualitative evidence for the formation of chemical bonds during the collisions. It should  be pointed out  here  that  the  number  and  form of the localization regions are  strongly dependent  on the choice of  the isovalue. An isovalue of ELF $=$ 0.8 has proven to be a useful standard for the classical valence compounds \cite{doi:10.1002/anie.199718081}.

\begin{figure}[htp]
  \centering
  \includegraphics[width=9cm,height=8cm]{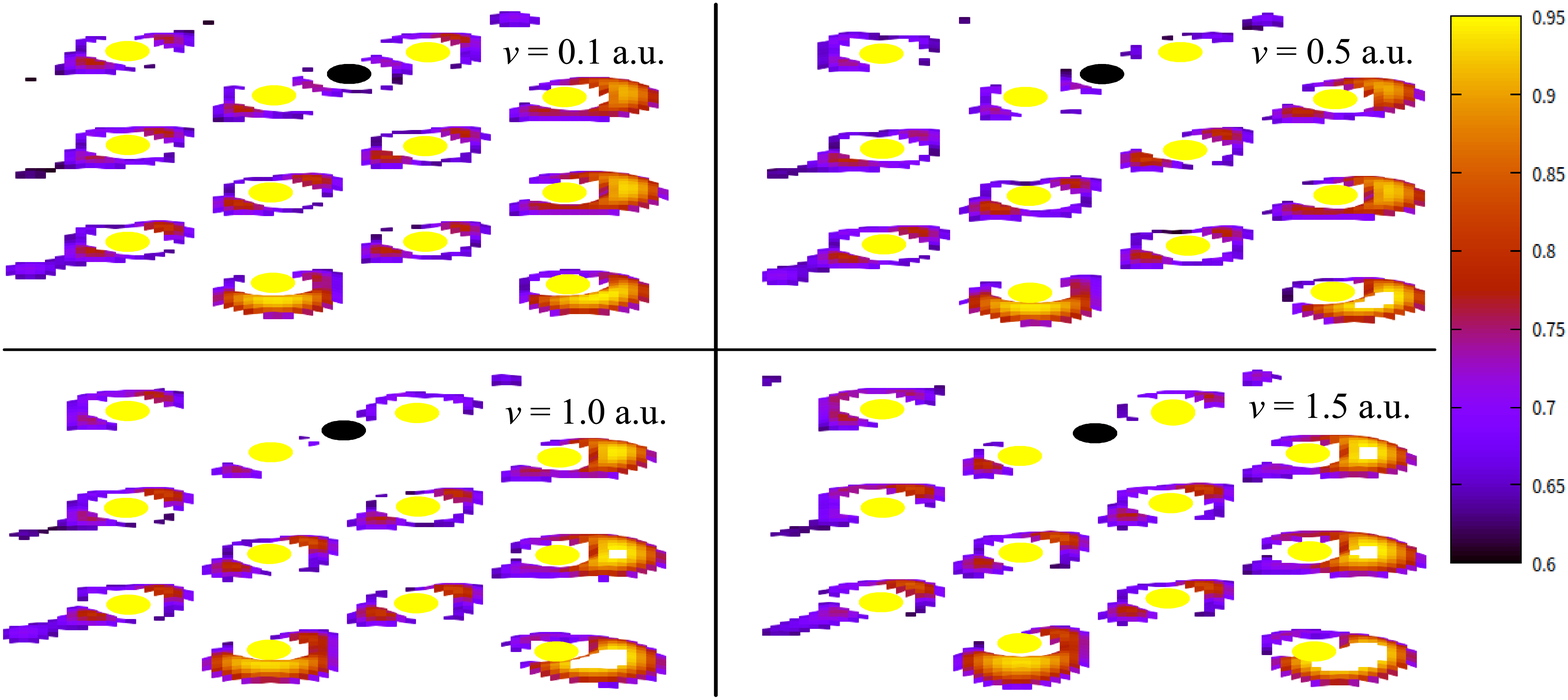}\\
  \caption{(Color online) Contours of the ELF on the plan y=-0.679. The Si$^{4+}$ ions is $on$ $the$ $fly$ with velocities of 0.1, 0.5, 1.0 and 1.5 a.u. respectively, the instantaneous positions of projectiles are at z = 4.69. The yellow circles show the position of host silicon atoms, the black circles show the position of projectiles.}
\end{figure}

The different \emph{$S_{e}$} characteristics between the high velocity regime and the lower velocity regime is studied from the strength of chemical bonds.  Figure~3 presents four contours including the time-dependent ELF on an $x$-$z$ plan at $y$ = -0.679 for the channeling of silicon ions with velocities of 0.1, 0.5, 1.0 and 1.5 a.u. at the same point inside the crystal. It is to be noted Fig.~3 depicts chemical bonds between ions on this plan with atoms outward/inward. There is an evident distribution of ELF with relative large values between the 0.1 a.u. silicon ion and the host atom. The ELF distribution becomes subtle with the increase of speed, and when the velocity reachs 1.5 a.u., there is no ELF distribution between the projectile and the host atom at all. This can be rationalized by the fact that it takes time for the excited electron captured by the projectile Si ion to form chemical bonds with neighboring host atoms, slow movement leaves more time for the excited electron to evolve and form the chemical bonds.

\begin{figure}[htp]
  \centering
  \includegraphics[width=8cm,height=6cm]{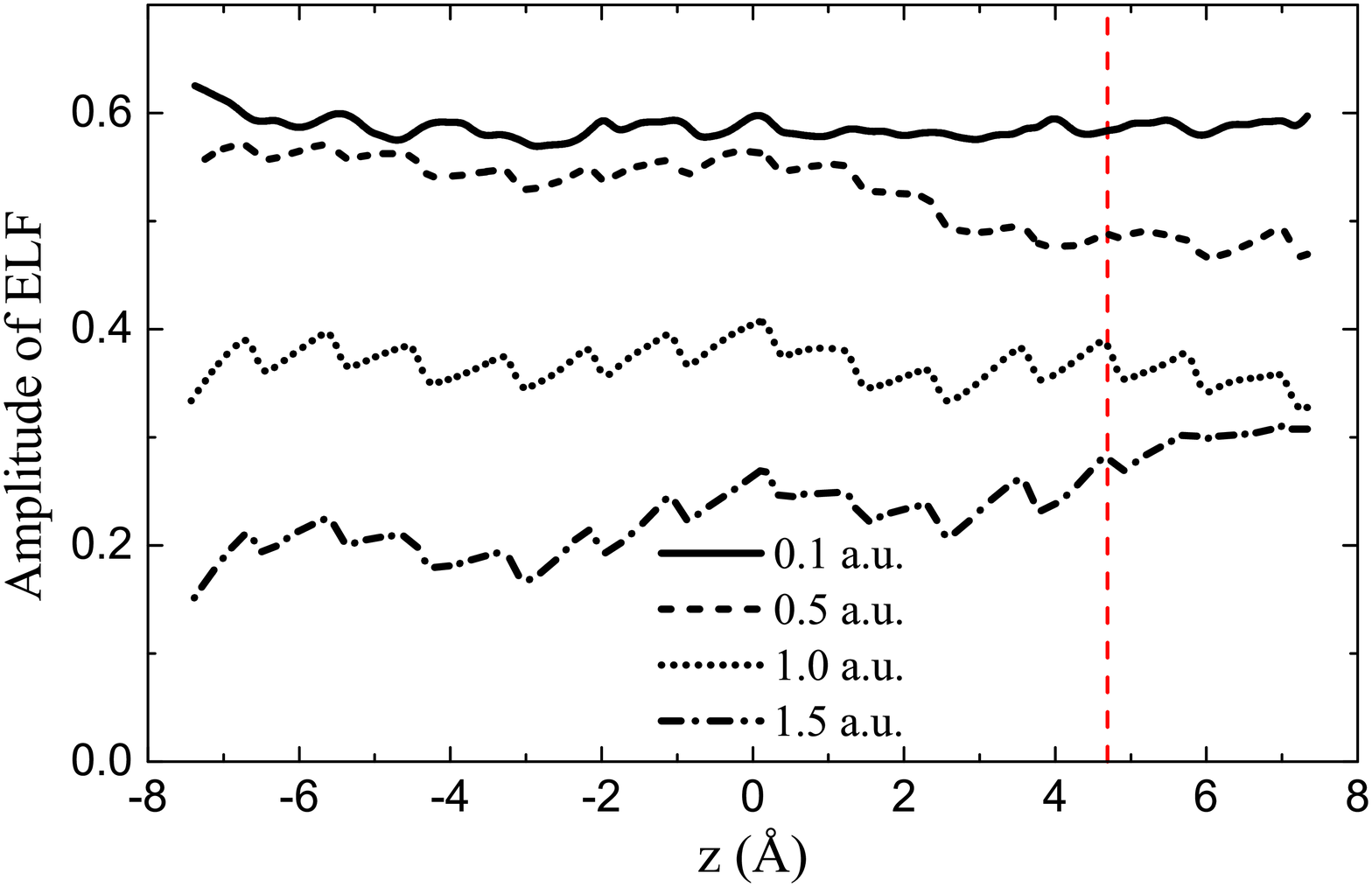}\\
 \caption{(Color online) Average ELF within a radius of 1.08 {\AA} around the projectiles with velocities of 0.1, 0.5, 1.0 and 1.5 a.u. respectively at each step along the whole trajectory.
  The vertical dashed line denotes to the projectile position of $z$ = 4.69 shown in Fig.~3.}
\end{figure}

The amplitude of ELF induced by the silicon ion in real time is quantified by averaging the ELF within a radius of 1.08 {\AA} around the projectile in the time--dependent calculation. This value is deemed as the index of chemical bonds strength between the projectile and the host atoms i.e. bond order. Higher ELF density around the projectile means more stable chemical bonds between the projectile and the host atom, and more kinetic energy of the projectiles will be dissipated to break the chemical bonds.

In Fig.~4, we present the amplitude of ELF induced by Si$^{4+}$ ions with different velocities moving through $<$100$>$ channel. The amplitude of ELF decreases with velocity, which is consistent with the trend of \emph{$S_{e}$} (Fig.~1), indicating that formation and breaking of chemical bonds contributes directly to energy loss. The trend of  \emph{$S_{e}$} in Fig.~1 turn steeper below the velocity 0.3 a.u., this can be rationalized  by the fact that the decrease of electron excitation reduce the effect of chemical force in this velocity regime, since the formation of chemical bond is dynamic of excited electrons.



Theoretical study from first principles the non-adiabatic interaction of slow silicon ions with bulk Si has been presented. 
A quantitative agreement between the experimental data and our results is achieved. For low energies the \emph{$S_{e}$} deviates from linear response theory when silicon ions are channeled along $<$100$>$ direction. This is interpreted as a consequence of additional short range chemical force experienced by the low velocity projectile. Our work has unveiled one of
the most fundamental consequences of the non-adiabaticity of the electron-nuclear system, namely the modification of electronic energy loss that results from the formation and breaking of chemical bond. These are surprising and interesting
new chemical aspects of ion-solid interaction. Chemical bonds is expected to be of significance in evaluating the stopping of slow heavy ions interaction problems.
We hope this work may stimulate further experimental and theoretical work on chemical interaction between the ions and the target electrons.

This work was supported by the National Natural Science Foundation of China under Grant Nos.\,51727901, 11635003, 11025524 and 11161130520, National Basic Research Program of China under Grant No.\,2010CB832903, and the European Commissions 7th Framework Programme (FP7-PEOPLE-2010-IRSES) under Grant Agreement Project No.\,269131.

%

\end{document}